%% file: arXiv.tex
\documentclass[sigconf]{acmart}

\renewcommand\footnotetextcopyrightpermission[1]{} 
\usepackage{hyperref}
\hypersetup{
	pdfauthor={...},
	pdftitle={...},
	pdfsubject={...},
	urlcolor=blue,
}
\usepackage{breakurl}

\usepackage{booktabs} 
\usepackage{todonotes}
\setcopyright{rightsretained}
\usepackage{float}

\begin{document}
\title{Towards Understanding the Evolution of Vocabulary Terms in\\ Knowledge Graphs}

\author{Mohammad Abdel-Qader}
\affiliation{%
  \institution{Kiel University and ZBW -- Leibniz Information Centre for Economics}
}
\email{stu120798@informatik.uni-kiel.de}

\author{Ansgar Scherp}
\affiliation{%
  \institution{ZBW -- Leibniz Information Centre for Economics and Kiel University}
}
\email{a.scherp@zbw.eu}

\renewcommand{\shortauthors}{M. Abdel-Qader and A. Scherp}

\begin{abstract}
	Vocabularies are used for modeling data in Knowledge Graphs (KG) like the Linked Open Data Cloud and Wikidata. During their lifetime, the vocabularies of the KGs are subject to changes. New terms are coined, while existing terms are modified or declared as deprecated. We first quantify the amount and frequency of changes in vocabularies. Subsequently, we investigate to which extend and when the changes are adopted in the evolution of the KGs. We conduct our experiments on three large-scale KGs for which time-stamped snapshots are available, namely the Billion Triples Challenge datasets, Dynamic Linked Data Observatory dataset, and Wikidata. Our results show that the change frequency of terms is rather low, but can have high impact when adopted on a large amount of distributed graph data on the web. Furthermore, not all coined terms are used and most of the deprecated terms are still used by data publishers. There are variations in the adoption time of terms coming from different vocabularies ranging from very fast (few days) to very slow (few years). Surprisingly, there are also adoptions we could observe even before the vocabulary changes are published. Understanding this adoption is important, since otherwise it may lead to wrong assumptions about the modeling status of data published on the web and may result in difficulties when querying the data from distributed sources.
	
\end{abstract}

%
%




\maketitle

\input{samplebody-conf}

\bibliographystyle{ACM-Reference-Format}

\newpage
\appendix
\section{Appendix}
\subsection{Adopting the Newly Created Terms}

In this section, we illustrate all the vocabularies with newly created terms and their adoption  over time.

\begin{figure}[H]
	\centering
	\includegraphics[width=1.0\columnwidth]{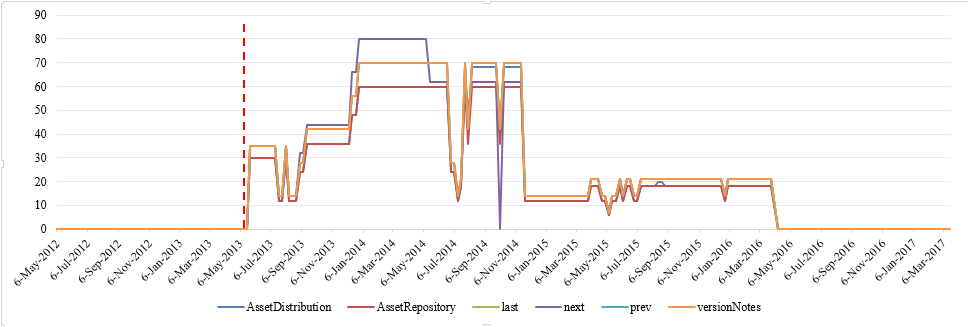}
	\caption{Adoption of terms from the Asset Description Metadata Schema (ADMS) vocabulary}
	\label{fig:ADMS}
\end{figure}

\begin{figure}[H]
	\centering
	\includegraphics[width=1.0\columnwidth]{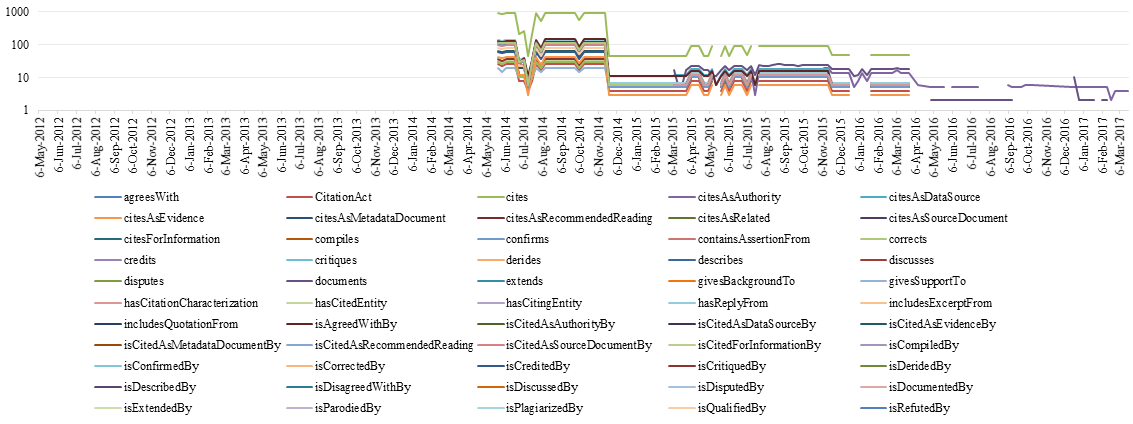}
	\caption{Adoption of terms from the Citation Typing Ontology (CiTO)}
	\label{fig:Cito}
\end{figure}

\begin{figure}[H]
	\centering
	\includegraphics[width=1.0\columnwidth]{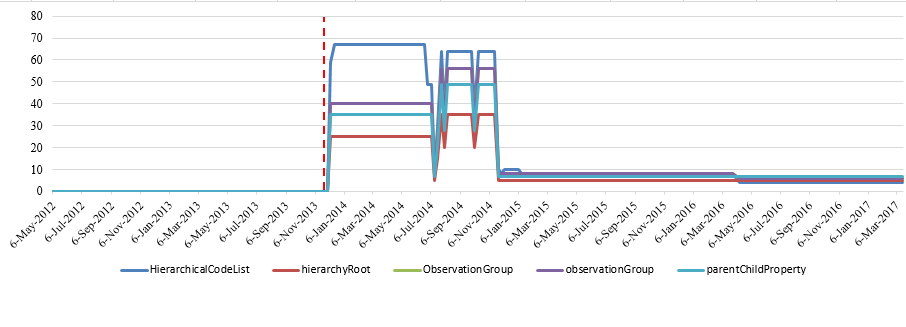}
	\caption{Adoption of terms from the data cube vocabulary (Cube)}
	\label{fig:Cube}
\end{figure}

\begin{figure}[H]
	\centering
	\includegraphics[width=1.0\columnwidth]{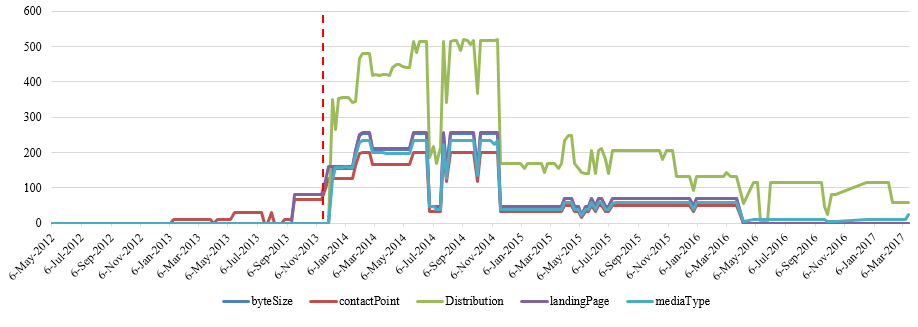}
	\caption{Adoption of terms from the Data Catalog Vocabulary (DCAT)}
	\label{fig:DCAT}
\end{figure}

\begin{figure}[H]
	\centering
	\includegraphics[width=1.0\columnwidth]{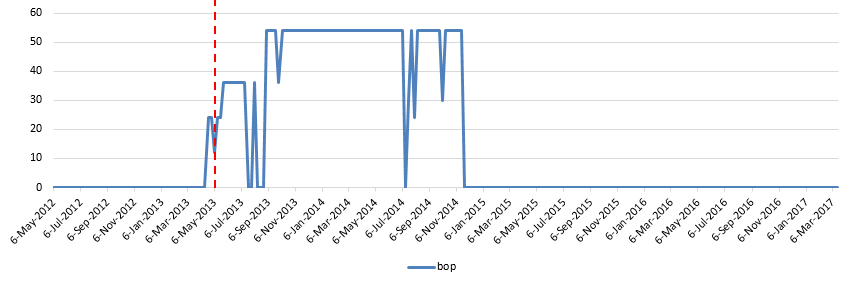}
	\caption{Adoption of terms from the vocabulary for jobs (emp)}
	\label{fig:emp}
\end{figure}

\begin{figure}[H]
	\centering
	\includegraphics[width=1.0\columnwidth]{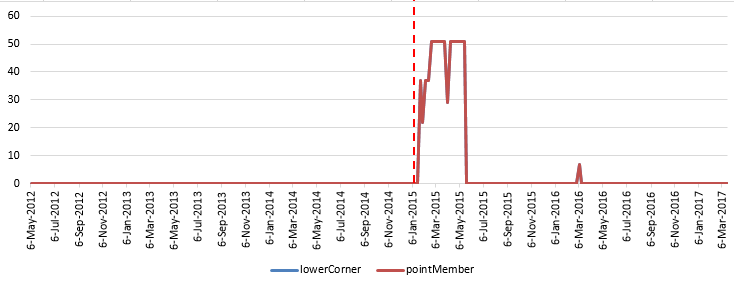}
	\caption{Adoption of terms from the ontology for geometry (geom)}
	\label{fig:geom}
\end{figure}

\begin{figure}[H]
	\centering
	\includegraphics[width=1.0\columnwidth]{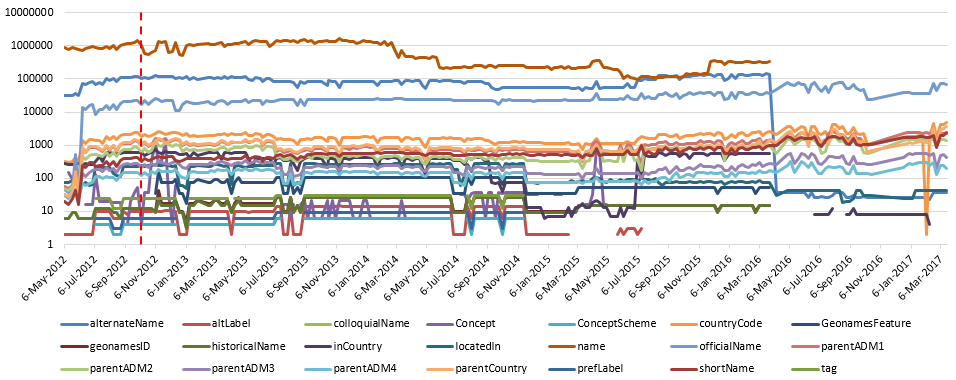}
	\caption{Adoption of terms from the Geonames ontology (GN)}
	\label{fig:GN}
\end{figure}

\begin{figure}[H]
	\centering
	\includegraphics[width=1.0\columnwidth]{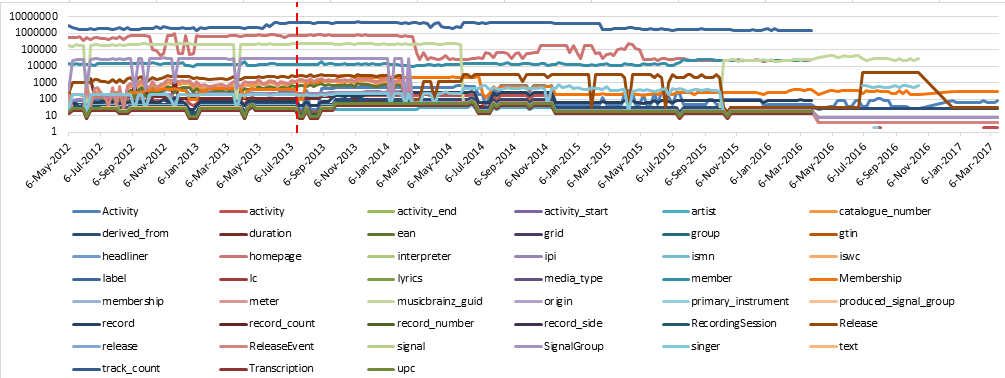}
	\caption{Adoption of terms from the music ontology (mo)}
	\label{fig:mo}
\end{figure}

\begin{figure}[H]
	\centering
	\includegraphics[width=1.0\columnwidth]{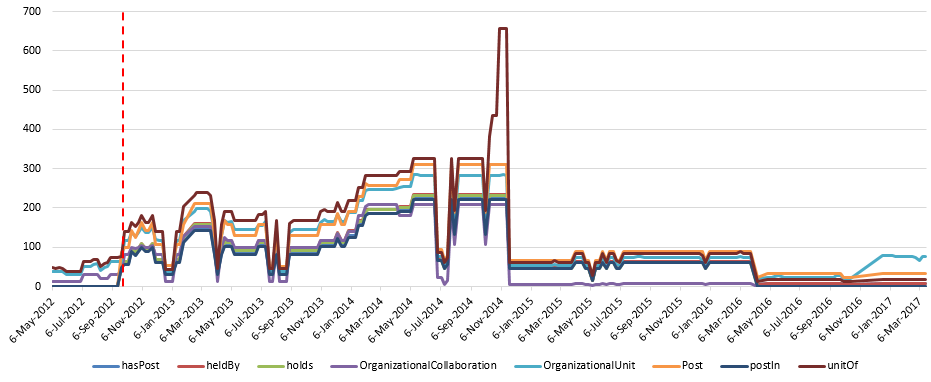}
	\caption{Adoption of terms from the Core organization ontology (org)}
	\label{fig:org}
\end{figure}

\begin{figure}[H]
	\centering
	\includegraphics[width=1.0\columnwidth]{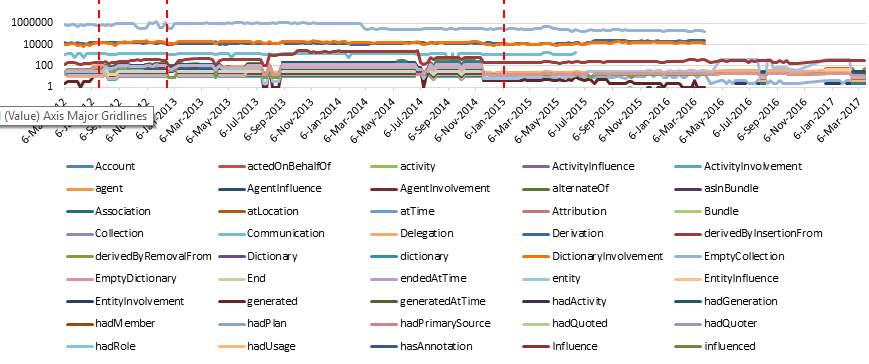}
	\caption{Adoption of terms from the W3C PROVenance Interchange (Prov) vocabulary}
	\label{fig:prov}
\end{figure}

\begin{figure}[H]
	\centering
	\includegraphics[width=1.0\columnwidth]{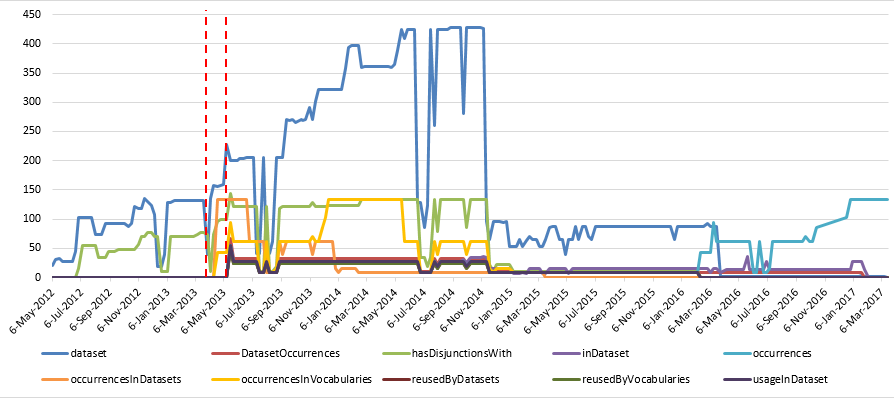}
	\caption{Adoption of terms from the Vocabulary of a Friend (Voaf)}
	\label{fig:Voaf}
\end{figure}

\end{document}

%% file: samplebody-conf.tex
\section{Introduction}

Vocabulary terms define the schema of Knowledge Graphs (KGs) such as the Linked Open Data (LOD) cloud or Wikidata.
After ontology engineers built and published the first version of a vocabulary, the terms are subject to changes to reflect new requirements or shifts in the domains the vocabularies model.
So far it is unknown how such vocabulary changes are reflected by the KGs that are using the terms.
For example, data providers may not be aware that changes on the vocabulary terms happened, since it occurs rather rarely~\cite{kafer2013observing}.
Explicitly triggering data providers to update their model is also challenging due to the distributed nature of KGs such as the LOD cloud.
However, in general data providers may be interested in being notified when certain vocabulary term changes happen.
But until today, data providers lack proper tools and services to track whether and what kind of changes on vocabulary terms happened.
Likewise, ontology engineers lack a tool that reflects the adoption status of data w.r.t. their vocabularies and changes on the vocabulary terms.

We present a methodology and its implementation in a service to improve the understanding of vocabulary changes and how they are adopted in evolving KGs.
Formally, we understand a vocabulary $V$ as set of terms $T$. 
A term $T$ is either a class $C$ or a property $P$.
A set of terms relates to a vocabulary as $T(V) = C(V) \bigcup P(V)$.
Changes on a vocabulary $V$ are changes on its terms, i.\,e., the classes and properties.
Data that uses classes and properties of a changed vocabulary should be updated according to those changes. 
This is needed especially, when a change is classified as deprecation. 
But there are also other types of changes like additions, modifications, or renaming a term.
In a previous work~\cite{abdel2016qualitative}, we manually conducted a qualitative analysis of vocabulary evolution on the LOD cloud. 
We analyze the changes for a set of vocabularies by clarifying which terms changed, the type of change, and if those changes were done on terms defined in the vocabularies or on the classes and properties that were imported from other vocabularies. Based on this qualitative analysis, we have developed a system to automatically extract changes on a set vocabularies and analyze the adoption time for the vocabulary changes, i.\,e., we monitor when changes on classes and properties are adopted by the KGs. 

In this paper, we are studying the evolution of vocabulary terms in KGs. Our first research question is, when are the newly created terms of vocabularies adopted in KGs? Second, what is the usage rate of classes and properties for a set of vocabularies in each dataset, and are the Pay Level Domains which that highly use existing and deprecated terms? Third, are the deprecated terms are still used in KGs?
%
We consider the two the basic types of changes, namely addition and deletion. Any other change like a modification can be expressed by the basic change types.
%
As datasets, we use three well-known KGs. 
First, we use the Dynamic Linked Data Observatory (DyLDO) dataset \cite{kafer2012towards}, which are weekly snapshots for a set of Linked Data documents from the LOD cloud.
Second, we analyze the Billion Triples Challenge (BTC)\footnote{\url{http://challenge.semanticweb.org/}, last accessed: June 29, 2017} datasets crawled from the LOD cloud in 2009, 2010, 2011, 2012, and 2014. 
From both datasets, we extract the Pay Level Domains (PLDs) that adopted changes of vocabulary terms. 
By PLD, we refer to any web domain that requires payment at a Top Level Domain (TLD) or country code TLD (cc-TLD) registrar~\cite{lee2009irlbot}. 
%
As third dataset, we analyze the evolution of the Wikidata\footnote{\url{https://www.wikidata.org}, last accessed: June 29, 2017} KG. 
We extract the vocabulary terms from Wikidata and determine if changes on the vocabulary were done (additions or deprecations) and how these changes were adopted. 
%
%
Our experiments show that even if the frequency of vocabularies terms changes is rather low, they have a large impact on the real data. Furthermore, we found that most of the newly coined terms are adopted in less than one week after their publishing date. However, there are terms that are only adopted after several months or up to a few years after the date of creation. Moreover, there are also some terms adopted before their official publishing date.
Additionally, many deprecated terms are still in use in KGs. Therefore we can claim that many of those terms are in practice not really deprecated. Finally, we found that the percentages of the actually unused terms in KGs are more than 50\% for most vocabularies, especially in the BTC dataset.
%
%
For Wikidata, we extract which terms that are introduced to the vocabulary are actually never being used in the Knowledge Graph and aim to explain why there are terms that are not adopted. 
We found no deprecation of terms found in the Wikidata vocabulary. 
Moreover, there are 17 terms not used because they are for defining properties and their types.
We think our work does help ontology engineers to select classes and properties that fit their needs by having a clear view about their adoption time and adoption rate, and help them in updating their own ontologies. 
In a subsequent step, we aim to provide this approach as freely available web service to track vocabulary changes and their adoption.

The remainder is structured as follows: 
In Section~\ref{RW}, we review the related work. 
Our methodology is presented in Section~\ref{Meth}, followed by a description of the datasets and vocabularies extracted from these datasets in Section~\ref{Dataset}.
Section~\ref{sec:results} describes the results of our experiments.
We discuss the findings and outcomes of our experiments in Section~\ref{sec:Discussion}, before we conclude.

\section{Related Work}
\label{RW}

In terms of analyzing the use of structured data on the web, Meusel et al. \cite{meusel2015web} analyzed the evolution and adoption of schema.org. 
They made a comparison of the usage of schema.org terms over four years by extracting the structured data of the web pages that use this vocabulary from \textit{WebDataCommon} Microdata datasets. They extracted only the quads whose  object or predicate contains \textit{schema.org}. They discovered that not all terms in schema.org are used and deprecated terms are still used, as it is also shown in this work. 
Furthermore, they found that publishing new types and properties is preferred over using schema.org's extension mechanism. 
Guha et al. \cite{guha2016schema} investigated the usage of the schema.org vocabulary in the structured data of a set of web pages. 
They analyzed a sample of 10 billion web pages crawled from \textit{Google index} and \textit{WebDataCommon}. 
They found that about 31\% of those pages had some \textit{schema.org} elements and estimated that around 12 million websites are using \textit{schema.org} terms. 
For their analyses, the changes on vocabulary terms were not considered as it is investigated in this work.
Furthermore, the authors said the wide usage of this vocabulary's terms was caused by the vast support from third-party tools, such as \textit{Drupal} and \textit{Wordpress}. 
Mihindukulasooriya et al. \cite{mihindukulasooriya2016collaborative} conducted a quantitative analysis for studying the evolution of DBpedia, Schema.org, PROV-O, and FOAF ontologies. They proposed some recommendations such as the need of dividing large ontologies into modules to avoid duplicates when adding new terms. Furthermore, they suggest adding  provenance information beside the generic metadata when the change occurred. 
Dividino et al. \cite{dividino2013change} analyzed how the usage of RDF classes and properties on the LOD cloud changed over time. 
Thus, they analyzed the combination of classes and properties that describe a resource but did not investigate whether a vocabulary and its terms have changed.
The authors applied their analysis on a dataset of 53 weekly snapshots from the DyLDO dataset, as it is also investigated in this work.
Furthermore, Dividino et al. \cite{dividino2015strategies} developed novel strategies for updating large-scale LOD datasets to keep local caches up-to-date. 
Two setups were investigated. 
The first setup was a single-step update of a local data cache, i.\,e., taking only one snapshot as history into account.
The second setup considered updates over a long period of time that used information about the total sum of data evolution as one of its features. 
The evaluation presented the performance for each setup based on bandwidth limitations. 
However, for both experiments, the changes on vocabulary terms were not considered as it is investigated in this work.
Vandenbussche et al. \cite{vandenbussche2017linked} published a report that describes Linked Open Vocabularies (LOV). 
The report provides statistics about using LOV and its capabilities such as the total number of classes and properties and the top-10 search terms using LOV (from January 2015 to June 2015). But the report does not include information about adopting new terms and from which PLDs.
Rathachari et al. \cite{chawuthai2016presenting} proposed a model that facilitates the understanding of organisms.  
Their model presents the changes in taxonomic knowledge in RDF form. 
The proposed model acts as a history tracking system for changing terms. But their model does not give information about how and when the terms are used, and which PLDs adopted changed terms. 
Schaible et al.~\cite{schaible2014survey} published a survey for the most preferred strategies for reusing vocabulary terms. 
The participants were 79 Linked Data experts and practitioners and were asked to rank several LOD modeling strategies. 
The survey concluded that terms that are widely used are considered as a better approach. 
Furthermore, vocabularies that are frequently used is a more important argument for reuse than the frequency of a single vocabulary term (ignoring the frequency of the vocabulary where the term belongs to). 
This survey can help to understand why there are some terms frequently used and why some of them are not used at all.
Over six months, K\"afer et al.\cite{kafer2013observing} observed the documents retrieved from the DyLDO dataset they crawled. 
They analyzed those documents using different factors, their lifespan, the availability of those documents and their change rate. 
Also, they analyzed the RDF content that is frequently changed (triple added or removed).
Finally, they observed how links between documents are evolved over time.
Their study is important for the communities in different areas such as smart caching, link maintenance, and versioning.
But their analyses does not include information about adopting new and deprecated terms.

\section{Analysis Method}
\label{Meth}
Our analysis method consists of two steps:
First, we determine vocabularies that have more than one published version on the web.
Second, we investigate how the changed terms of vocabularies are adopted and used in the evolving KGs.
For the first step, we based on Schmachtenberg et al. \cite{schmachtenberg2014adoption} who published a report with detailed statistics about a large-scale snapshot the LOD cloud. 
The snapshot comprises seed URIs from the datahub.io dataset\footnote{\url{http://datahub.io/group/lodcloud}}, the BTC 2012 dataset\footnote{\url{http://km.aifb.kit.edu/projects/btc-2012/}}, and the public-lod@w3.org mailing list\footnote{\url{http://lists.w3.org/Archives/Public/public-lod/}}.
We select a set of vocabularies that satisfy the following set of conditions and characteristics:
	(1) The vocabulary should have at least two versions published on the web to make a comparison between them.
	(2) We check if these two versions are covered by the dataset we investigate.
	For example, for the DyLDO dataset there is to be one version of the vocabularies that has been published after May 6th, 2012.
	This is needed, since at this date the first snapshot of the DyLDO dataset has been crawled. 
	(3) The vocabulary terms are to be directly used for modeling some data, i.\,e., there needs to be a direct use of the vocabulary terms by at least one triple in the published dataset.
	In contrast, vocabularies could also be just linked from a data publisher, where changes of external vocabularies may not have any impact on the published data.

Based on the criteria above, we examined 134 of the most used vocabularies listed in the state of the LOD cloud 2014 report by Schmachtenberg et al. \cite{schmachtenberg2014adoption}.
We found 18 vocabularies that have more than one version. 
From them, 13 vocabularies have changes (additions or deprecations) on terms created by the ontology engineers of those vocabularies in the timeframe of the considered datasets. 
We downloaded the different versions of the extracted vocabularies using the Linked Open Vocabularies (LOV) observatory\footnote{\url{http://lov.okfn.org/dataset/lov}, last accessed: June 20, 2017}. LOV is using a script that automatically checks for vocabulary changes in a daily basis and stores the detected version locally.
We extracted the changes between each two successive versions of a vocabulary by using Prot\'eg\'e 4.3.0, which is a free, open source ontology editor application that allows users to create, edit, and compare ontologies\footnote{\url{http://protege.stanford.edu}, last accessed: June 29, 2017}. 
%

Subsequently, we investigate in the second step how changed vocabulary terms are used in the evolving KGs. 
We extract all PLDs from the crawled triples that use the terms from the 13 vocabularies above.
We use \textit{Guava}\footnote{\url{https://github.com/google/guava/}, last accessed: June 29, 2017} library version 16.0.1 to extract the PLD from URLs. 
We record, besides the date of the first appearance of a vocabulary term, also the number of triples that contain the term. 
This information is then used to compute the adoption time of vocabulary term changes over the dataset snapshots.

\section{Datasets and Vocabularies}\label{Dataset}
\subsection{Datasets}
We apply our analysis approach on three large-scale KGs.
The first two are DyLDO and BTC and are obtained from the Linked Open Data cloud, and the third is Wikidata.
Below, we briefly characterize the datasets.

The Dynamic Linked Data Observatory (DyLDO) is a repository to store weekly snapshots from a subset of web data documents~\cite{kafer2012towards}. 
They started crawling snapshots since May 2012. For our study, we parse 242 snapshots (from May 2012 until March 2017). 
%
The Billion Triple Challenge (BTC) is part of the Semantic Web Challenge\footnote{\url{http://challenge.semanticweb.org/}, last accessed: July 10, 2017}, with a main goal to crawl a dataset from the LOD cloud. 
We used all available BTC datasets which were crawled in the years 2009, 2010, 2011, 2012 and 2014 to analyze the adoption of the extracted vocabularies in our study.

Wikidata is a knowledge base to collaboratively store and edit structured data. 
It is a free and multilingual repository\footnote{\url{https://www.wikidata.org/}, last accessed: June 29, 2017}.
In order to analyze the Wikidata vocabulary, we first extract the terms introduced by this vocabulary. 
Using the RDF Exports from Wikidata page\footnote{\url{http://tools.wmflabs.org/wikidata-exports/rdf/exports.html}, last accessed: June 29, 2017}, we parse the terms and properties from the RDF dump files that were generated using the Wikidata toolkit\footnote{\url{https://github.com/Wikidata/Wikidata-Toolkit}, last accessed: June 29, 2017}. 
We assume that the first snapshot of those files is the first version of the Wikidata vocabulary, and based on this assumption we parse the next dump files to extract the changes to the first version, and so on.
Overall, there are 25 RDF dump files (from  April 2014 until August 2016). 
Using those files, we extract the terms that are added or deprecated. 
Subsequently, we parse the Wikidata statements RDF dump files to extract the adoption of terms to analyze the adoption behavior for the Wikidata vocabulary's terms.

\subsection{Extracted Vocabularies}
The vocabularies that fulfill the conditions for the automated analysis of vocabulary terms evolution are listed in Table \ref{table:vocabs}. The table also shows the number of versions of each vocabulary that are used in this study. Furthermore, it shows the total number of changes (additions and deletions) occurred during the vocabularies evolution.

\begin{table} [H]
	\centering
	\caption{Overview of the vocabularies and their number of changes. The \# changes column represents the total number of additions and deletions counted during the vocabulary evolution.}
	\label{table:vocabs}
	\begin{tabular}{lcc}
		\toprule
		Vocabulary & \# versions & \#  changes\\
		\midrule
		Asset Description Metadata Schema \\ (ADMS)\footnote{\url{https://www.w3.org/TR/vocab-adms/}, last accessed: June 10, 2017} & 2 &18\\
		Citation Typing Ontology (CiTO)\footnote{\url{http://www.sparontologies.net/ontologies/cito/source.html}, last accessed: June 10, 2017} & 3 &218\\
		The data cube vocabulary (Cube)\footnote{\url{http://www.w3.org/TR/vocab-data-cube/}, last accessed: June 10, 2017} & 2&6 \\
		Data Catalog Vocabulary (DCAT)\footnote{\url{https://www.w3.org/TR/vocab-dcat/}, last accessed: June 10, 2017} & 2 &13\\
		A vocabulary for jobs (emp)\footnote{\url{http://lov.okfn.org/dataset/lov/vocabs/emp}, last accessed: June 10, 2017} & 2 &1\\
		Ontology for geometry (geom)\footnote{\url{http://data.ign.fr/def/geometrie/20160628.htm}, last accessed: June 10, 2017} & 2& 2\\
		The Geonames ontology (GN)\footnote{\url{http://www.geonames.org/ontology/documentation.html}, last accessed: June 10, 2017} & 7 &31\\
		The music ontology (mo)\footnote{\url{http://www.geonames.org/ontology/documentation.html}, last accessed: June 10, 2017} & 2 &46\\
		Open Annotation Data Model (oa)\footnote{\url{http://www.openannotation.org/spec/core/}, last accessed: June 10, 2017} & 2 &31\\
		Core organization ontology (org)\footnote{\url{https://www.w3.org/TR/vocab-org/}, last accessed: June 10, 2017} & 2& 8\\
		W3C PROVenance Interchange (Prov)\footnote{\url{https://www.w3.org/TR/prov-o/}, last accessed: June 10, 2017} & 5 &168\\
		Vocabulary of a Friend (voaf)\footnote{\url{http://lov.okfn.org/vocommons/voaf/v2.3/}, last accessed: June 10, 2017} & 4 &8\\
		An extension of SKOS for representation \\ of nomenclatures (xkos)\footnote{\url{http://rdf-vocabulary.ddialliance.org/xkos.html}, last accessed: June 10, 2017} & 2&1\\
		\bottomrule
	\end{tabular}
\end{table}

\section{Results}\label{sec:results}
In Section \ref{sec:results}, we summarize our findings after parsing the DyLDO and BTC datasets and extracting the required information for conducting this study. 
Section \ref{DyLDOandBTC} lists our findings on changes of the 13 vocabularies in terms of additions and deprecations.
Furthermore, we summarizes the results of usage and adoption of the 13 vocabularies in KGs datasets. 
Section \ref{wikiChanges} summarizes the findings that are related to the Wikidata vocabulary in terms of changes, usage, and adoption.

\subsection{Changes, Usage, and Adoption Time of Terms on the LOD Cloud}
\label{DyLDOandBTC}
\begin{figure*}
	\centering
	\includegraphics[height=3.5in, width=7in]{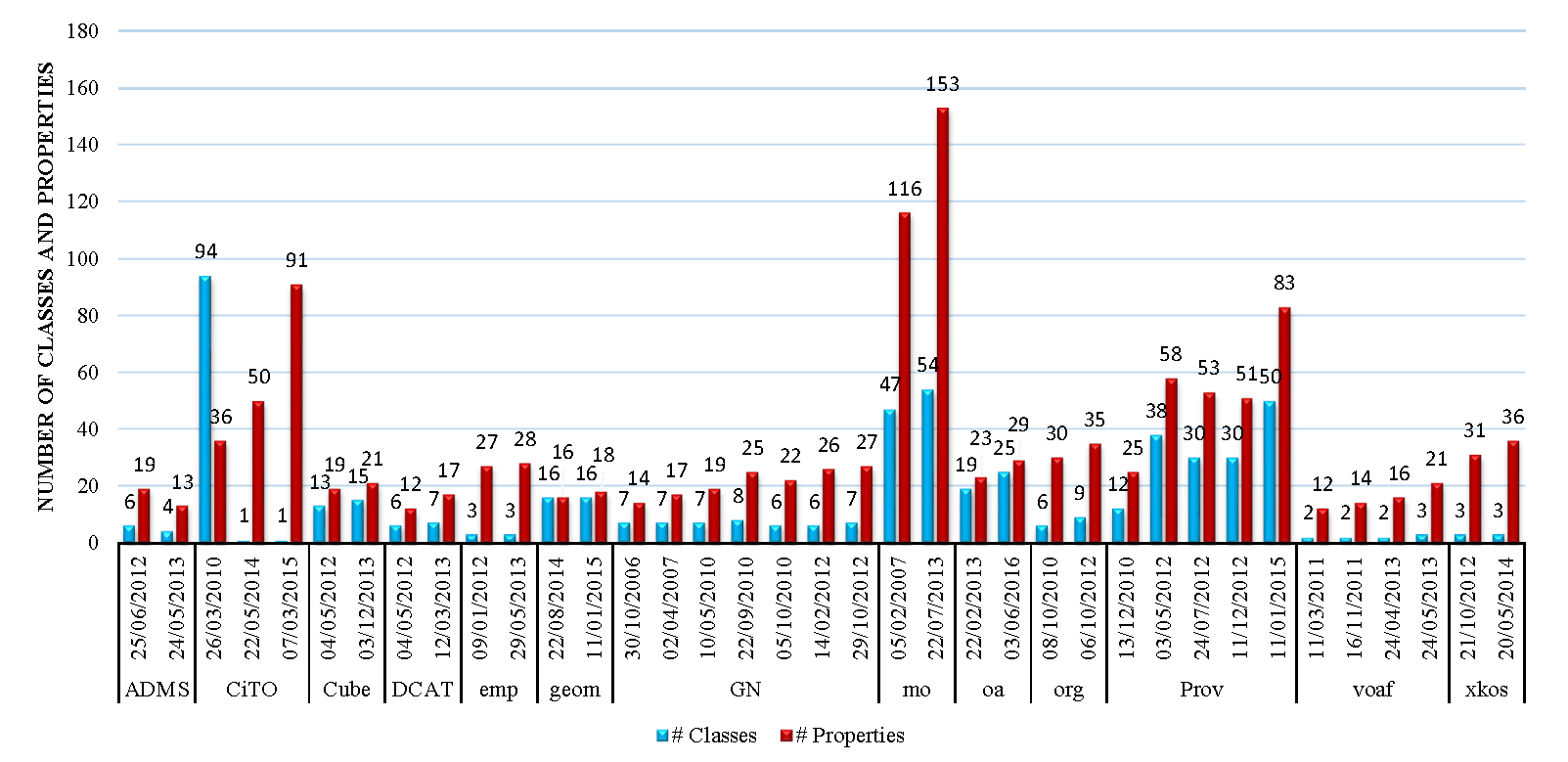}
	\caption{The total number of classes and properties for each of the 13 vocabularies versions in our study. The blue bar (left) represents the number of classes and the red bar (right) represents the number of properties.}
	\label{fig:fig1}
\end{figure*}

Figure \ref{fig:fig1} shows the 13 vocabularies in our study and the total number of classes and properties for each version of those vocabularies. Please note that most of them have an increased number of terms except two vocabularies. The number of classes and properties is decreased for the \textit{ADMS} vocabulary, and \textit{CiTO} had a huge decline in the number of classes.

During our analysis, we noticed that some of the deprecated properties are recreated again after they were deprecated. 
These recreated terms related to the \textit{CiTO} and \textit{GN} vocabularies. 
The \textit{CiTO} vocabulary deprecated 18 properties in May 2014 (they were introduced in March 2010), and then they were recreated again in the version that was published in March 2015. The \textit{GN} vocabulary was recreated three deprecated properties: \textit{alternateName} (creation: October 2006, deprecation: September 2010, recreation: February 2012), \textit{name} (creation: October 2006, deprecation: September 2010, recreation: October 2010), and \textit{shortName} (creation: September 2010, deprecation: May 2010, recreation: February 2012). We can notice that \textit{CiTO} has recreated those properties after around ten months of deprecation. Furthermore, for the \textit{GN} vocabulary, two out of three deprecated terms are recreated in the new version after about 17 months, and the remaining one was recreated fast (after 13 days).



For each dataset, Table \ref{table:table6} shows the PLDs (based on all PLDs extracted during our analysis) that used terms related to the 13 vocabularies. 
For most vocabularies, we notice that there are no variety in the PLDs that use terms from the 13 vocabularies, i.\,e., there is one PLD that has the highest usage rate of terms. Furthermore, we can observe that \textit{org} and \textit{Cube} vocabularies are used by many PLDs.
Table \ref{table:NewTerms} shows an analysis for adopting the newly created terms for each vocabulary in our study. Additionally, it shows the total number of instances that the newly created terms appeared in. Furthermore, the table shows the mean ($\mu$) and the standard deviation ($\sigma$) values in terms of days for adopting the newly created terms. We can notice that most of the newly coined terms are adopted, in less than 10 days. Furthermore, we notice that all new terms related to the \textit{oa} and \textit{xkos} vocabularies are not adopted at all.
Table \ref{table:table8} shows the number of triples use terms from the 13 vocabularies. We can observe that \textit{geom} does not appear in all the BTC datasets. Additionally, we can say that \textit{emp} and \textit{oa} vocabularies almost did not appear in the BTC and DyLDO datasets.
Table \ref{table:table9} shows the PLDs with a high usage rate terms from the 13 vocabularies. We can notice that \textit{geonames.org} has a high usage rate in the BTC2009 and BTC2010 datasets, while \textit{dbtune.org} has a high usage rate in the BTC2014 and DyLDO datasets.

\begin{table}
	\small
	\caption{A list of PLDs with the highest usage rate (based on all PLDs extracted during our analysis) that highly used terms related to the 13 vocabularies with changes. The (Amount) represent the total numbers of triples that the corresponding PLD appears in.}
	\label{table:table6}	
	\begin{tabular}{ccc}
		
		\toprule
		Vocabulary & BTC (Amount) & DyLDO (Amount)\\
		\midrule 
		ADMS & w3.org (5K) & w3.org (253K)\\
		\hline
		CiTO & \vtop{\hbox{\strut ontologycentral.com} \hbox{\strut (3.3M)}}  &\vtop{\hbox{\strut ontologycentral.com} \hbox{\strut (4.7M)}} \\
		\hline
		Cube & \vtop{\hbox{\strut ontologycentral.com} \hbox{\strut (1.2M)}} & \vtop{\hbox{\strut ontologycentral.com} \hbox{\strut (1.8M)}}\\
		\hline
		DCAT & \vtop{\hbox{\strut ontologycentral.com} \hbox{\strut (5M)}}& \vtop{\hbox{\strut ontologycentral.com} \hbox{\strut (4.8M)}}\\
		\hline
		emp & purl.org (11K) & purl.org (14K)\\
		\hline
		geom & - & ign.fr (5.5K)\\
		\hline
		GN & geonames.org (90M)& geonames.org (40M)\\
		\hline
		mo &dbtune.org (84M) & dbtune.org (76M)\\
		\hline
		oa & w3.org (7K)& w3.org (13K)\\
		\hline
		org &data.gov.uk (21K)&w3.org (111K)\\
		\hline
		Prov & w3.org (62K) & dbpedia.org (1M)\\
		\hline
		voaf & purl.org (381K) & purl.org (9K)\\
		\hline
		xkos & 270a.info (609K) & 270a.info(4K)\\
		\bottomrule
	\end{tabular}
\end{table}

\begin{table}
	\caption{The adoption of newly created terms for each of the vocabularies in this study. The \% of used terms column represent the percentage of the actually used terms compared to the total number of new created terms, and the \# instances column represents the total amount of triples the newly created terms appeared in.}
	\label{table:NewTerms}
	\begin{tabular}{ccccc}
		\toprule 
		Vocabulary &  \% of used terms& \# instances & $\mu$ (days) & $\sigma$ (days)\\
		\midrule 
		ADMS & 100\% & 31K &  7 & 0\\
		CiTO & 100\% & 281K& 7 & 0\\ 
		Cube & 100\% & 15K & 7 & 0\\ 
		DCAT & 100\% &104K&  8.4 & 3.13\\	 
		emp & 100\% & 4K& 7 & -\\	 
		geom & 100\% &16K&  420 & 0\\	 
		GN & 100\% &160M&  127.76 & 255.33\\	 
		mo & 100\% &45M&  8.75 & 9.68\\	 
		oa & 0\% & - & -\\	 
		org & 100\% &173K&  7 & 0\\	 
		Prov & 85\% &121M&  30.15 & 37.49\\	 
		voaf & 90\% & 75K& 43.33 & 68.58\\	 
		xkos & 0\% &  - & -\\
		\bottomrule 
	\end{tabular}
\end{table}

\begin{table}
	\small
	\caption{Number of triples in the DyLDO and BTC datasets that use terms from the 13 vocabularies.}
	\label{table:table8}	
	\begin{tabular}{ccccccc}
		\toprule
		Vocabulary & BTC09 & BTC10 & BTC11 & BTC12 & BTC14 & DyLDO\\
		\midrule 
		ADMS & 12 & 2 & 4 & 26 & 7K & 337K
		\\
		
		CiTO & 0 & 0 & 4 & 4.5K & 305K& 1M
		\\
		
		Cube & 0 & 0 & 40K & 8.4K & 56M & 12M
		\\
		
		DCAT & 0 & 0 & 12 & 4.5K & 317K & 9.8M
		\\
		
		emp & 0 & 0 & 0 & 0 & 238 & 26K
		\\
		
		geom & 0 & 0 & 0 & 0 & 0 & 5.7K
		\\
		
		GN & 81M & 7.4M & 477K & 441K & 1M& 55M
		\\
		
		mo & 2M & 1.7M & 12M & 4M & 102M& 83M
		\\
		
		oa & 0 & 0 & 0 & 0 & 192& 23K
		\\
		
		org & 0 & 9 & 129 & 11K & 20K& 700K
		\\
		
		Prov & 0 & 9 & 43 & 902 & 3M& 4M
		\\
		
		voaf & 0 & 0 & 2 & 0 & 4K& 1M
		\\
		
		xkos & 0 & 0 & 0 & 0 & 610K & 194K
		\\
		\bottomrule
	\end{tabular}
\end{table}

\begin{table}
	\centering
	\caption{List of PLDs which have the highest usage rate (based on all PLDs extracted during our analysis) that used terms from the 13 vocabularies. The Repetition column represents the number of triples that the PLDs appears in.}
	\label{table:table9}
	\begin{tabular}{ccc}
		\toprule
		Dataset & PLD & Repetition\\
		\midrule 
		BTC2009 & geonames.org & 81M\\
		
		BTC2010 & geonames.org & 7M\\
		
		BTC2011 & zitgist.com & 2.6M\\
		
		BTC2012 & rdfize.com & 3.8M\\
		
		BTC2014 & dbtune.org & 81.5M\\
		
		DyLDO & dbtune.org & 160M\\
		
		\bottomrule
	\end{tabular}
\end{table}

We noticed that most of the deprecated terms are still used after they were marked as deprecated. We found that \textit{geonames.org} is the PLD which have the highest adoption rate for deprecated terms. 
The number of deprecated terms in all the 13 vocabularies and used by the \textit{geonames.org} PLD, and extracted during our analysis of DyLDO, BTC2011, BTC2012, and BTC2014 datasets are 6, 6, 3, and 49, respectively.

Table \ref{table:table11} shows the percentage of unused terms from vocabularies depending on the extracted triples from the BTC and DyLDO datasets. We can notice that all terms of the \textit{geom} vocabulary did not appear in BTC, beside most of the terms of \textit{oa, emp}, and \textit{CiTO}. In DyLDO, \textit{CiTO} has the highest percentage of unused terms.
Based on the triples extracted from DyLDO dataset, all the 37 created terms by \textit{Cube} vocabulary appeared in the extracted triples. \textit{org} vocabulary created 44 terms, and only 5 of them not appeared in the crawled triples of BTC dataset.

\begin{table}
	\small
	\caption{Number and percentage of unused terms per vocabulary for the BTC and DyLDO datasets. The \#terms column represents the total number of terms in the vocabulary.}
	\label{table:table11}
	\begin{tabular}{cccccc}
		\toprule
		Vocabulary & \# terms & BTC & DyLDO & BTC\% & DyLDO\%\\
		\midrule 
		ADMS & 31 & 21 & 1 & 68\% & 3\%\\
		CiTO & 220 & 158 & 132 & 72\% & 60\%\\
		Cube & 37 & 13 & 0 & 35\% & 0\%\\
		DCAT & 23 & 11 & 2 & 48\% & 9\%\\
		emp & 31 & 27 & 2 & 87\% & 6\%\\
		geom & 34 & 34 & 1 & 100\% & 3\%\\
		GN & 43 & 11 & 4 & 26\% & 9\%\\
		mo & 208 & 74 & 4 & 36\% & 2\%\\
		oa & 63 & 52 & 22 & 83\% & 35\%\\ 
		org & 44 & 9 & 5 & 20\% & 11\%\\ 
		Prov & 143 & 32 & 34 & 22\% & 24\%\\ 
		voaf & 24 & 8 & 2 & 33\% & 8\%\\
		xkos & 35 & 22 & 5 & 63\% & 14\%\\
		\bottomrule
	\end{tabular}
\end{table} 

\subsection{Changes, Usage, and Adoption Time of Terms in Wikidata}
\label{wikiChanges}
\begin{figure}
	\centering
	\includegraphics[width=1.0\columnwidth]{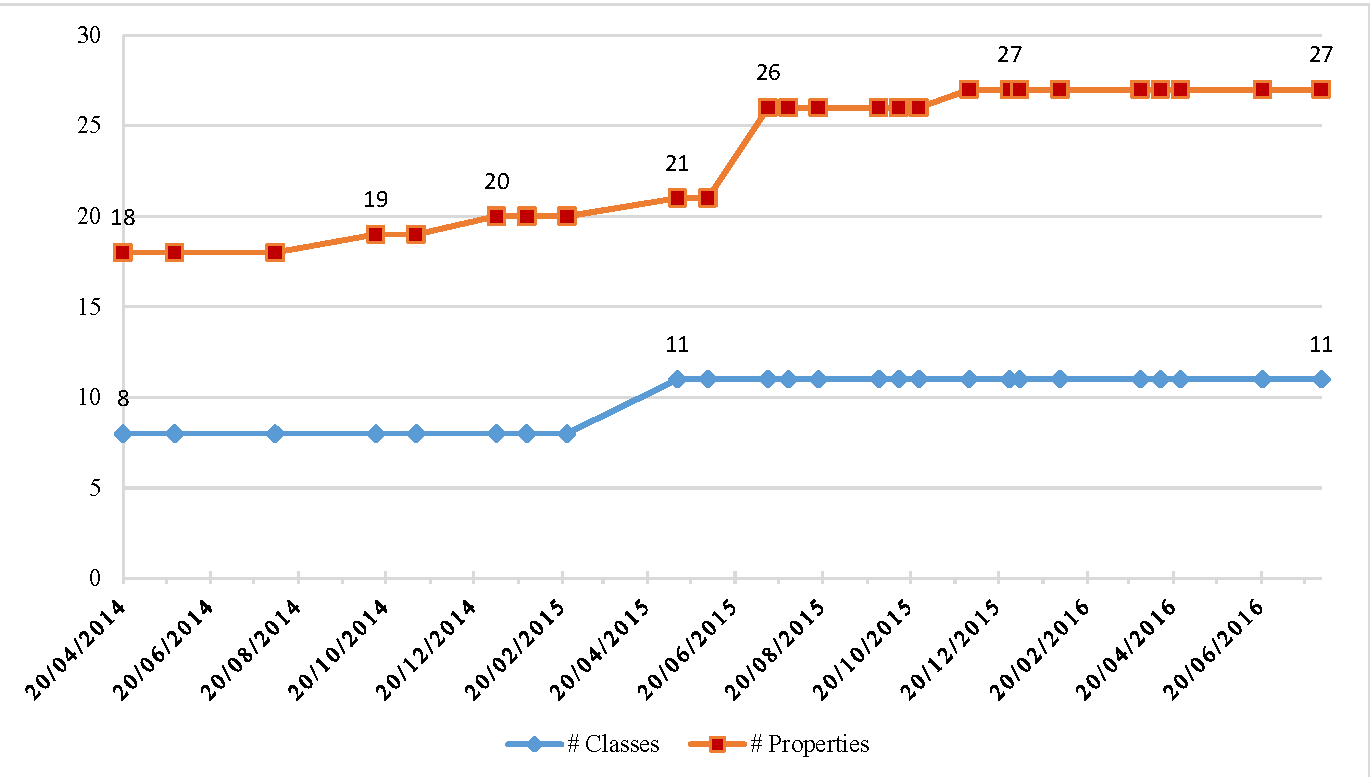}
	\caption{The total number of classes and properties in each RDF dump file generated by the Wikidata toolkit.}
	\label{fig:wiki}
\end{figure}

After parsing the terms and properties from the RDF dump files for the period from April 2014 until August 2016, we extract the added and deprecated terms. Figure \ref{fig:wiki} shows the total number of classes and properties in each Wikidata snapshot. 
One can see, it grows to reach 11 classes and 27 properties in August 2017.
Another important note is that there are no terms that are deprecated during the ontology evolution.

For the Wikidata vocabulary, ontology engineers added 3 classes and 9 properties during the analyzed period. The new classes are: \textit{DeprecatedRank, PreferredRank,} and \textit{NormalRank}. And the properties are: \textit{propertyTypeMonolingualText, propertyTypeProperty, rank, propertyQualifierLinkage, propertyReferenceLinkage, propertySimpleClaim, propertyStatementLinkage, propertyValueLinkage,} and \textit{quantityUnit}.

\begin{figure}
	\centering
	\includegraphics[width=1.0\columnwidth]{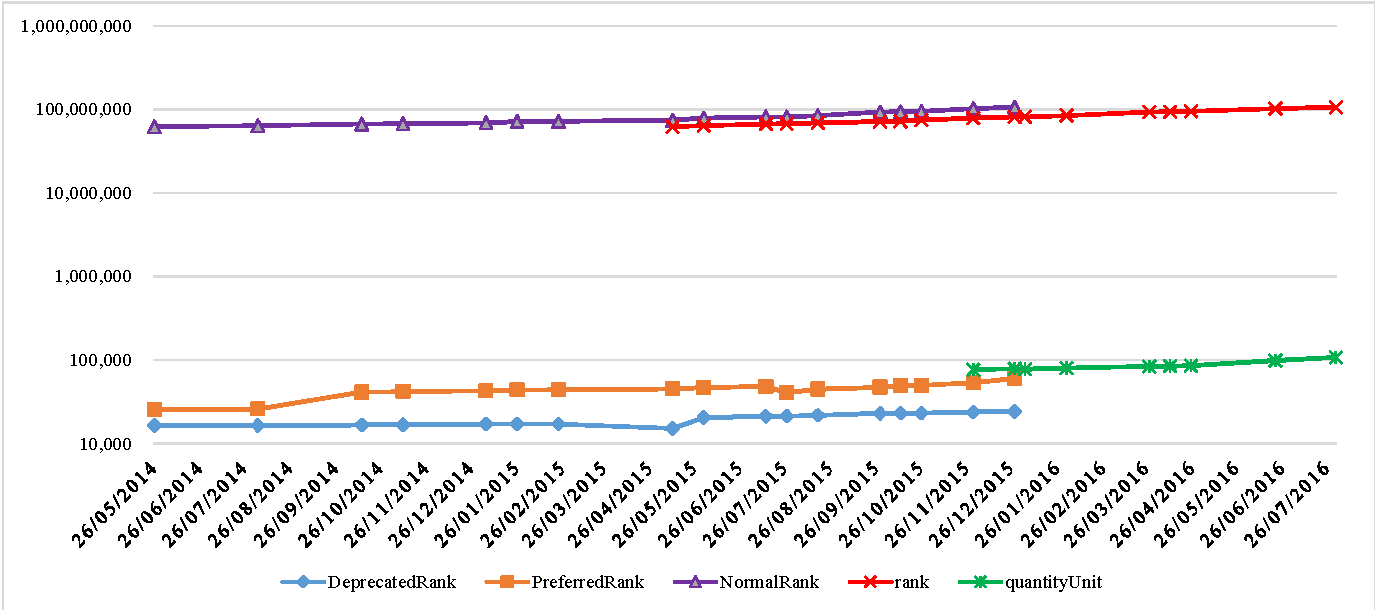}
	\caption{The usage amount in terms of triples for the newly created classes and properties on Wikidata vocabulary after parsing Wikidata RDF dump files.}
	\label{fig:wikiNew}
\end{figure}

Figure \ref{fig:wikiNew} shows the usage amount (in terms of triples) for newly created classes and properties on Wikidata. 
One can notice that the figure contains only 5 terms, while the actual number of newly coined terms is 12. This is because of there are 7 newly created terms are never used in Wikidata statements. Furthermore, we can notice that there is one class (\textit{NormalRank}) and one property (\textit{rank}) that have the highest usage amount of triples.
Another interesting point is that the actually used terms from the newly created ones is adopted directly after their creation date. This note also can be applied to the other terms of the Wikidata vocabulary.

\section{Discussion}\label{sec:Discussion}
In Section \ref{changes} we discuss the changes of terms of the 13 vocabularies in LOD datasets. Furthermore, we discuss the usage and adoption of the newly created terms. The Wikidata terms changes and adoption is discussed in Section \ref{Wiki-disc}.

\subsection{Changes, Usage, and Adoption Time of Terms on the LOD Cloud} \label{changes}

The number of additions and deprecations of terms are small. 
This is in line with existing studies\cite{abdel2016qualitative, meusel2015web,guha2016schema}.
There is one exception, the \textit{CiTO} vocabulary's first version, which is published on March 2010, consists of 94 classes and 36 properties. After around four years of being online, the ontology engineers published an updated version that consisted of only one class and 50 properties.  They removed all the 94 classes and replaced them with the new class \textit{CitationAct}. Furthermore, most of the 36 properties of the first version are removed and replaced with new ones. The latest version of \textit{CiTO} was published in March 2015. The number of classes did not change, but the number of properties increased to 91, and 18 of them were deprecated from the first version.
We can say that the ontology engineers almost built a new ontology compared with the first published version, because they removed most of the terms and replaced them with new ones.
This is important to note, since \textit{CiTO} has grown much in popularity (e.\,g., the BTC 2014 dataset contained over 300K triples compared to 40K triples in BTC 2011).
Another observation is that the \textit{ADMS} vocabulary is the only vocabulary that has a decline in their terms (for both classes and properties). 




For most vocabularies, we found that only one PLD has the highest usage rate of their terms, and there is no variety on the PLDs that use and adopt the vocabulary terms.
In \textit{ADMS}, \textit{w3.org} was the PLD that highly used terms belonging to \textit{ADMS}, but later (in 2015 and 2016) \textit{deri.de} used the most of \textit{ADMS} terms. On the other hand, there are some vocabularies that have been used from different PLDs. For example, \textit{Cube} has been highly used from \textit{ontologyCenter.com, esd.org.uk, linked-statistics.org}, and \textit{linkedu.en}. 

\begin{figure*}
	\centering
	\includegraphics[height=3.5in, width=7in]{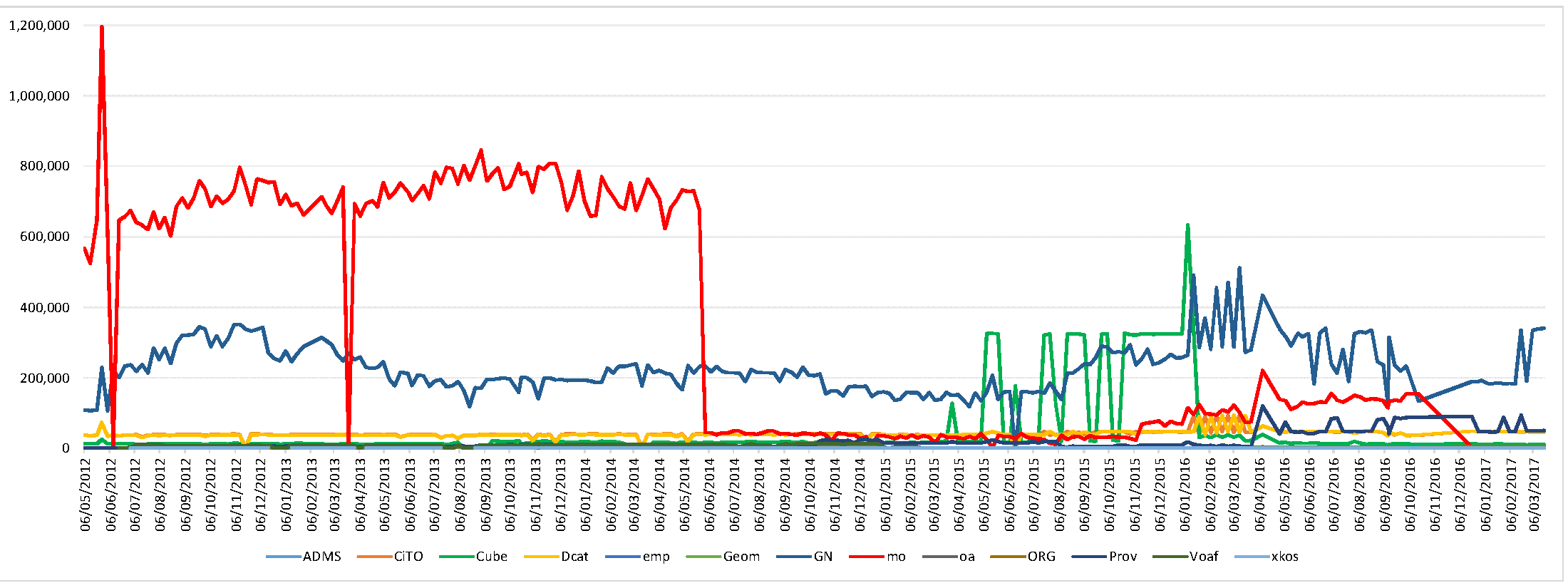}
	\caption{Amount of triples that had terms from the 13 vocabularies, in the DyLDO datasets. The X-axis represents the dates of DyLDO snapshots and Y-axis represents the total number of triples that use the terms of certain vocabularies.}
	\label{fig:dy}
\end{figure*}

Analyzing the DyLDO dataset, the amount of usage remains in the same range of triples in most of the vocabularies except for the \textit{mo} vocabulary (red line) and \textit{Cube} vocabulary (light green line) as shown in Figure \ref{fig:dy}, which shows the amount of triples for each vocabulary. The triples are calculated on a weekly basis. In our opinion, these variations are depending on the aim of vocabularies, the generality, and the clearance of terms.  
\textit{geonames.org} and \textit{dbtune.org} are the PLDs (based on all PLDs extracted during our analysis) that use terms from all the 13 vocabularies. 
In the BTC2009 and BTC2010 datasets, \textit{geonames.org} was the PLD that has a high usage rate of vocabulary terms. This is caused by the high number of triples using the \textit{GN} vocabulary in those years. Furthermore, \textit{dbtune.org} was the PLD with high usage rate in the BTC2014 and DyLDO snapshots from 2012 to 2014. 

\begin{figure}
	\centering
	\includegraphics[width=1.0\columnwidth]{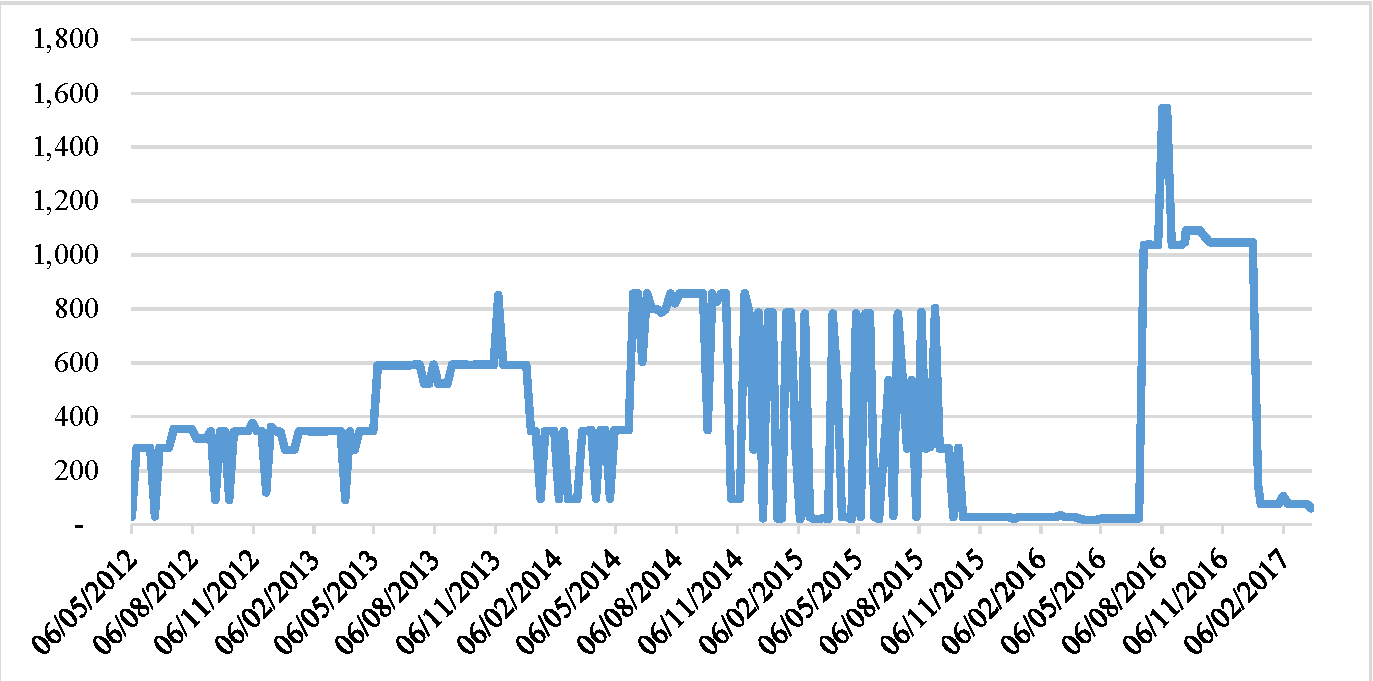}
	\caption{The usage of the class \textit{gn:Country} in the DyLDO dataset after the class deprecated in September 2010.}
	\label{fig:country}
\end{figure}

Although \textit{geonames.org} was the PLD which have the highest usage rate for the 13 vocabularies, it is also the PLD that has the most uses deprecated terms. For example, in the BTC2011 dataset, \textit{geonames.org} used six deprecated terms in about 522K triples. That number declined to three terms and about 181K triples in the BTC2012 dataset. But suddenly, in the BTC2014 the number of used deprecated terms has increased to be 49 terms, but in just about 5.5K triples. We assume that this was caused by not updating the documents and ontologies that used those deprecated terms. Figure \ref{fig:country} shows the usage of the term \textit{gn:Country} in the DyLDO dataset which deprecated in September 2010. The figure shows that it had an increased usage rate until August 2015, then it declined and increased again to reach the peak point of usage in August 2016.

Another interesting point is that some newly created terms are never adopted. For example, ontology engineers published a new version from \textit{oa} vocabulary in June 2016 that has 21 new classes and properties. None of those terms are adopted (at least until April 2017, which is the last DyLDO snapshot we parsed). While the first version of \textit{oa} is published in February 2013 with 42 terms, and all of them (except one term) are adopted in less than 3 months. We suggest that ontology engineers revise those unused terms. 

Most of the newly coined terms are adopted directly (in less than one week). It is interesting that we found some terms adopted before their official publishing date. We can conclude that some of the new versions of vocabularies are unofficially online and can be used before the official publishing announcement. We assume that in some cases, it may take time to finish the official procedures to publish the new version of the vocabulary.

\begin{figure}
	\centering
	\includegraphics[width=1.0\columnwidth]{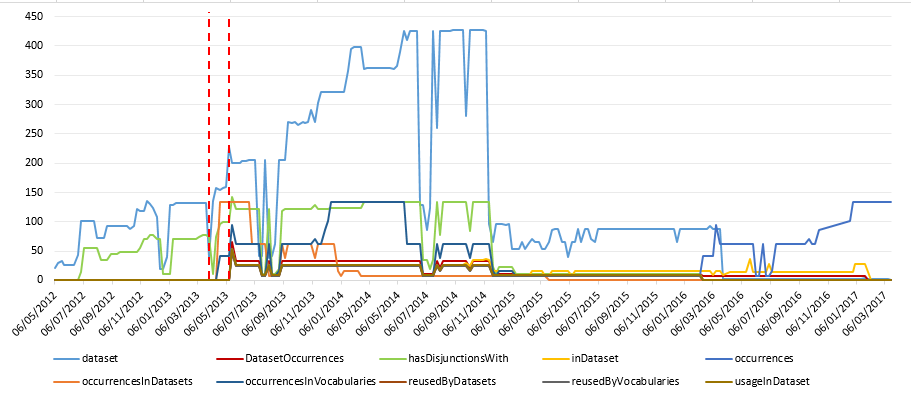}
	\caption{Amount of triples that use the \textit{voaf} vocabulary in the DyLDO datasets. The vertical red dashed lines represent the time of publishing new versions for \textit{voaf}.}
	\label{fig:voaf}
\end{figure}

The adoption of newly created terms are vary between vocabularies.
For example, the \textit{voaf} vocabulary has published four versions in March 2011, November 2011, April 2013, and May 2013. By parsing the DyLDO dataset, Figure \ref{fig:voaf} shows the usage amount in terms of triples for the new created terms only (which starts from November 2011 version). The \textit{occurrenceInVocabularies} property (yellow line) is created in April 2013, and appears in 42 triples in the same month of its creation. The amount of usage goes in an increasing rate to reach its peak point of usage in May 2014. We think adding or removing terms and the increased number of available ontologies, have an impact on using those vocabularies. For more information about all the 13 vocabularies in this study, please refer to the Appendix.

Furthermore, the average number of days shows that most of the terms have fast adoption time, but in some vocabularies the mean value were more than 120 days, such as in \textit{GN}. This average does not reflect the actual adoption behavior. From the new version of \textit{GN} which newly created 21 terms, 17 terms are adopted within 7 days, and the remaining 4 terms are adopted in 600 and 650 days. Those four terms influenced in such result.

\begin{figure}
	\centering
	\includegraphics[width=1.0\columnwidth]{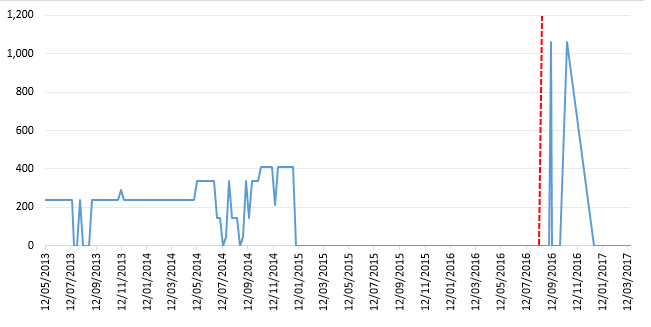}
	\caption{Amount of triples that use the \textit{oa} vocabulary in the DyLDO datasets. The vertical red dashed lines represent the time of publishing new versions for \textit{oa}.}
	\label{fig:oa}
\end{figure}

Publishing new versions of a vocabulary may influence the usage of terms. 
For example, Figure \ref{fig:oa} shows the usage of \textit{oa} vocabulary in the DyLDO dataset and when the new versions are published in June 2016. Even if the number of triples at its peak point (about 1K) maybe considered small, we can still notice an influence of these changes on the number of triples using the \textit{oa} vocabulary. Thus, we assume that vocabulary changes such as adding or removing terms is a cause for updating data in KGs.

\subsection{Changes, Usage, and Adoption Time of Terms in Wikidata} \label{Wiki-disc}
For the Wikidata vocabulary, an interesting point is that there are no deprecations of terms, even if they are not used during the time-frame investigated in this paper. Even if the terms are not used, ontology engineers does not mark them as deprecated (i.\,e., \textit{Article} class). We found that the other unused terms are created to define properties and their types.

For the Wikidata vocabulary, like most of the LOD vocabularies, there is a small number of additions (3 classes and 9 properties), and no terms are deprecated. Beside that, there is a huge difference in the amount of triples that terms appeared in. For example, \textit{NormalRank} and \textit{Statement} classes have being used in about 106M and 81M triples, respectively. While the other classes (except the \textit{Item} class) are used in less than 2.4M triples. The same observation can be made for properties, all of them appeared in less than 2.7M triples except the \textit{rank} property, which introduced in May 2015 with appearance in about 62M triples, and in August 2016 the usage raised up to reach about 106M triples.

Another interesting point is that  three classes (\textit{DeprecatedRank}, \textit{PreferredRank}, and \textit{NormalRank}) suddenly disappeared from Wikidata statements after the snapshot in December 2015, and after about 8 months of usage (creation date was in May 2015).
Another point is that there are two classes (\textit{Article} and \textit{Property}) and 15 properties which are not used in any statements of Wikidata. 
These are used to define the properties and their type (except the class \textit{Article}).

\section{Conclusion}
Even small changes of vocabulary terms can have a deep impact on the real data that use those terms. For instance, \textit{Prov} vocabulary remains in the range below than 25K triples before publishing its latest version in January 2015. After this version, the usage reached the level of 120K triples in 2016 and 100K in 2017.
 It is not surprising that deprecated terms are still in use. 
 However, surprisingly some deprecated terms have been recreated after some time by their vocabularies. 
 For instance, \textit{CiTO} has recreated 19 terms after they were marked to be deprecated. 
 Another example is \textit{gn}, which recreated 3 deprecated terms in its latest versions.